\documentclass[12pt]{iopart}
\usepackage{graphicx}

\begin{document}

\title{The thermal model on the verge of the ultimate test: particle production in Pb-Pb collisions at the LHC}

\author{A Andronic$^1$, P Braun-Munzinger$^{1,2,3}$, K Redlich$^{4}$, 
J Stachel$^5$}
\address{$^1$Research Division and ExtreMe Matter Institute EMMI and , GSI 
Helmholtzzentrum f\"ur Schwerionenforschung, Darmstadt, Germany
$^2$~Technical University, Darmstadt, Germany
$^3$~Frankfurt Institute for Advanced Studies, J.W. Goethe University,
Frankfurt, Germany,
$^4$~Institute of Theoretical Physics, University of Wroc\l aw, Poland,
$^5$~Physikalisches Institut, University of Heidelberg, Germany}

\begin{abstract}
We investigate the production of hadrons in nuclear collisions within the 
framework of the thermal (or statistical hadronization) model. 
We discuss both the ligh-quark hadrons as well as charmonium and provide 
predictions for the LHC energy. 
Even as its exact magnitude is dependent on the charm 
production cross section, not yet measured in Pb-Pb collisions,
we can confidently predict that at the LHC the nuclear modification factor of 
charmonium as a function of centrality is larger than that observed at RHIC
and compare the experimental results to these predictions.
\end{abstract}

One of the major goals of ultrarelativistic nuclear collision studies is to
obtain information on the QCD phase diagram \cite{pbm_wambach}. 
The experimental approach is the investigation of hadron production at chemical
freeze-out \cite{pbm_js}.  
Hadron yields measured in central heavy ion collisions from AGS up to RHIC 
energies can be described very well (see \cite{aat} and refs. therein) within 
a hadro-chemical equilibrium model.  In our approach the only parameters are 
the chemical freeze-out temperature $T$, the baryo-chemical potential $\mu_b$ 
and the fireball volume $V$.
The main result of these investigations was that the extracted temperature
values rise rather sharply from low energies on towards $\sqrt{s_{NN}}\simeq$10 GeV
and reach afterwards constant values near $T$=160 MeV, while the baryochemical
potential  decreases smoothly as a function of energy.
The limiting temperature \cite{hagedorn85} behavior suggests a connection to 
the phase boundary and it was, indeed, argued \cite{wetterich} that the 
quark-hadron phase transition drives the equilibration dynamically, at least 
for SPS energies and above. 
The conjecture of the tricritical point \cite{tri} was put 
forward for the lower energies.
Alternative conjectures are that the thermodynamical state 
is a generic fingerprint of hadronization \cite{stock,heinz} or is a feature 
of the excited QCD vacuum \cite{castorina}.

\begin{figure}[htb]
\begin{tabular}{lr} \begin{minipage}{.5\textwidth}
\includegraphics[width=.9\textwidth]{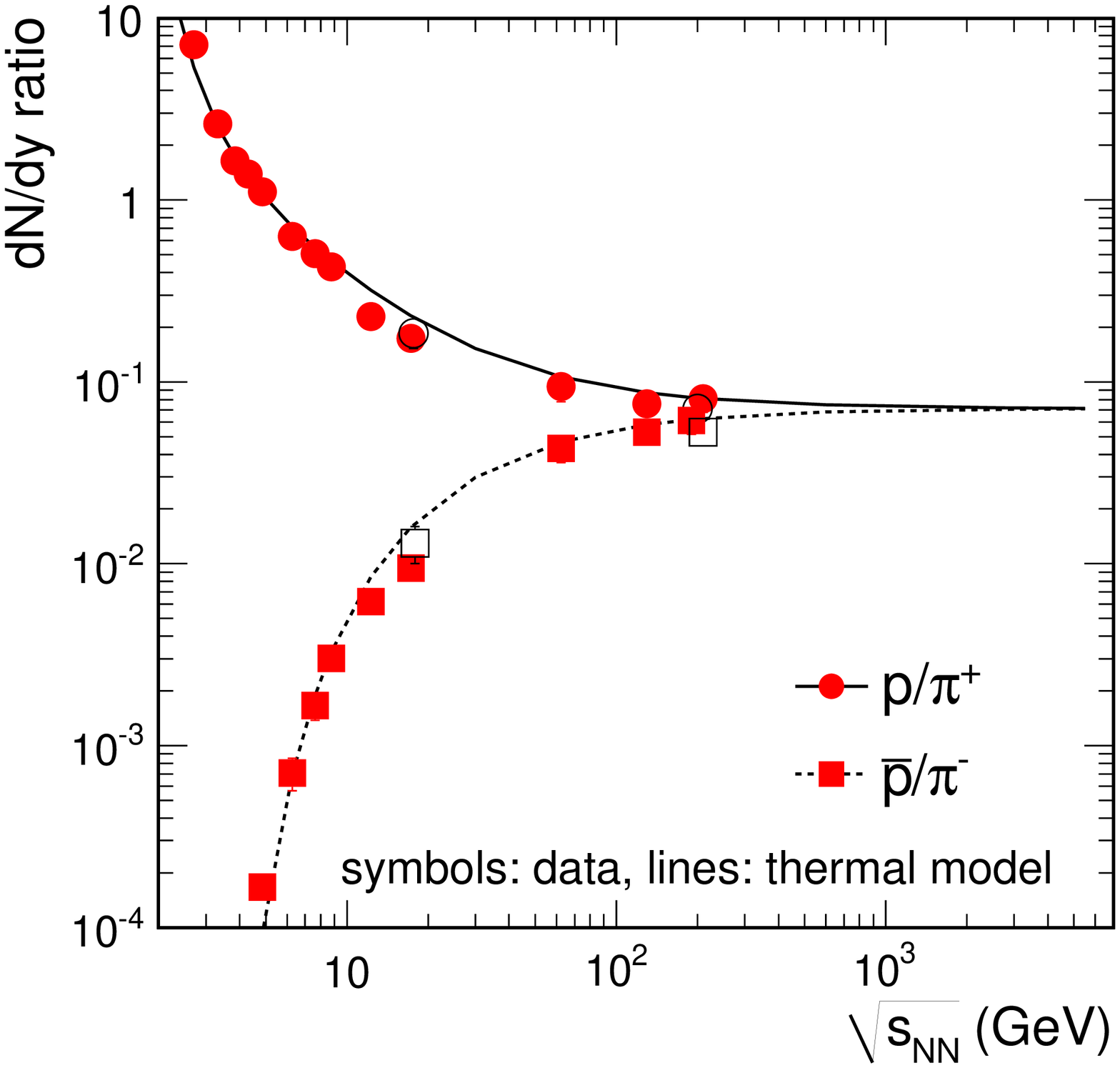}
\end{minipage} & \begin{minipage}{.5\textwidth}
\includegraphics[width=.9\textwidth]{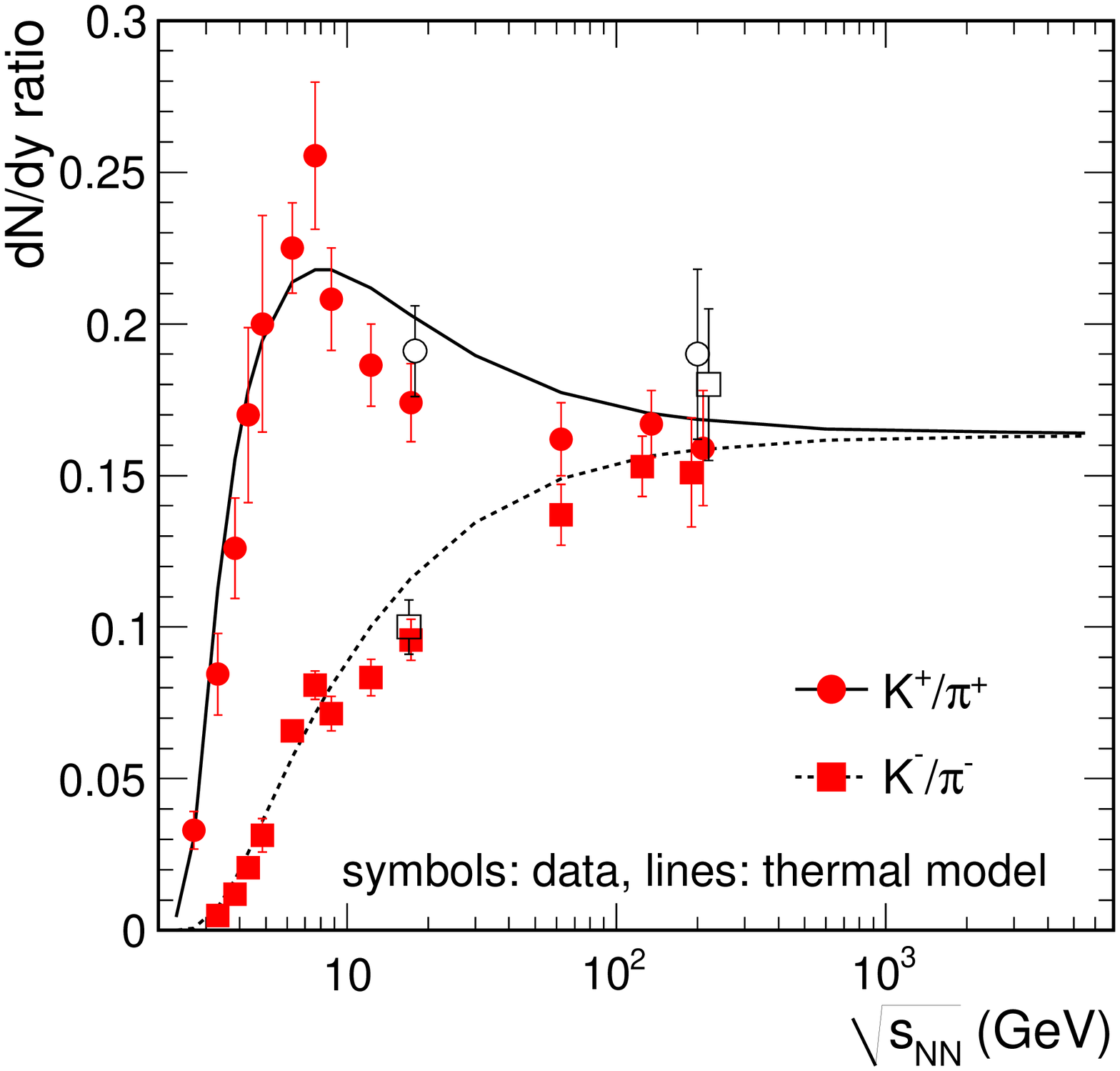}
\end{minipage}\end{tabular}
\caption{Energy dependence of the relative production ratios of protons ($\bar{p}$)
and kaons ($K^-$) to pions ($\pi^-$). The open symbols represent the data points
for the NA44 and PHENIX experiments at SPS and RHIC, respectively (the full 
symbols are for NA49 and STAR data). Note that ad-hoc feed-down subtraction was 
applied for STAR data for protons (25\%) and for PHENIX data for pions (10\%).
The errors of the $p/\pi$ and $\bar{p}/\pi$ data are smaller than the symbols.
}\label{fig_k2pi}
\end{figure}

The values of $T$ and $\mu_b$ obtained from fits can be parametrized as a function 
of $\sqrt{s_{NN}}$ with the expressions:
$T \mathrm{}=T_{lim}/[1+\exp(2.60-\ln(\sqrt{s_{NN}(\mathrm{GeV})})/0.45)]$
and
$\mu_b \mathrm{[MeV]}=1303/[1+0.286\sqrt{s_{NN}(\mathrm{GeV})}]$,
with the "limiting" temperature $T_{lim}$=164 MeV. 
The $\mu_b$ value expected at the LHC is around 1 MeV.
We employ these parametrizations to compare the model to data over a broad energy 
range. As an illustration, the production yields of protons and kaons relative 
to pions are shown in Fig.~\ref{fig_k2pi}, demonstrating that the model describes 
the data well (although smaller  $p/\pi^+$ and $\bar{p}/\pi^-$ ratios 
are measured by PHENIX and at SPS \cite{aat}).
The trends seen in the $p/\pi^+$ and $\bar{p}/\pi^-$ ratios reflect both the 
strong increase followed by saturation for $T$ and the strong decrease of 
$\mu_b$ as a function of $\sqrt{s_{NN}}$.
While the $K^-/\pi^-$ ratio shows a monotonic increase and saturation as a 
function of energy, the $K^+/\pi^+$ ratio shows a maximum at a beam energy of 
30 AGeV. In the thermal model this maximum occurs naturally at as an effect of the 
steep rise and saturation of $T$ and the strong monotonous decrease in $\mu_b$
\cite{aat}. 
\begin{figure}[htb]
\begin{tabular}{lr} \begin{minipage}{.5\textwidth}
\includegraphics[width=.9\textwidth]{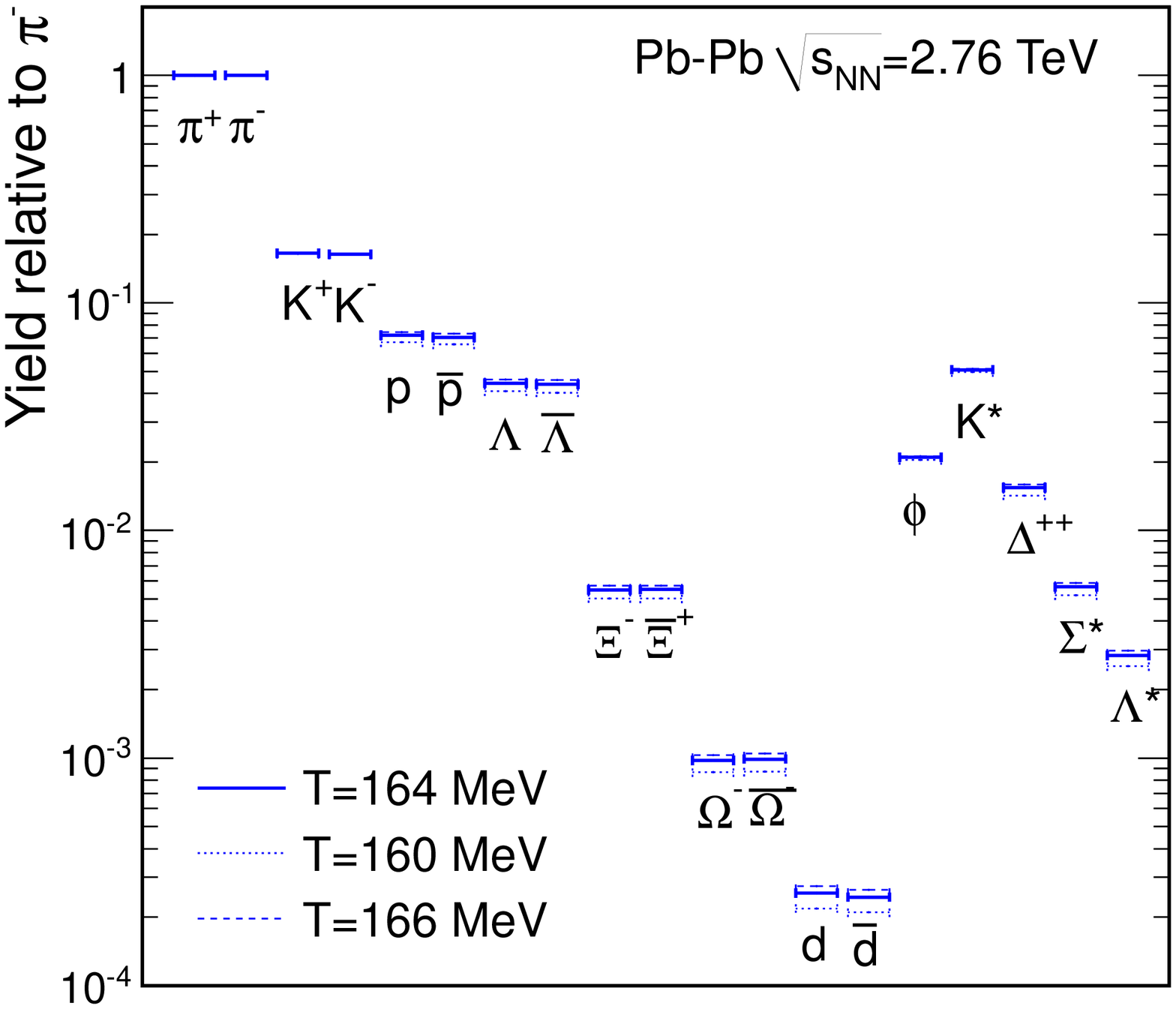}
\end{minipage} & \begin{minipage}{.5\textwidth}
\includegraphics[width=.9\textwidth]{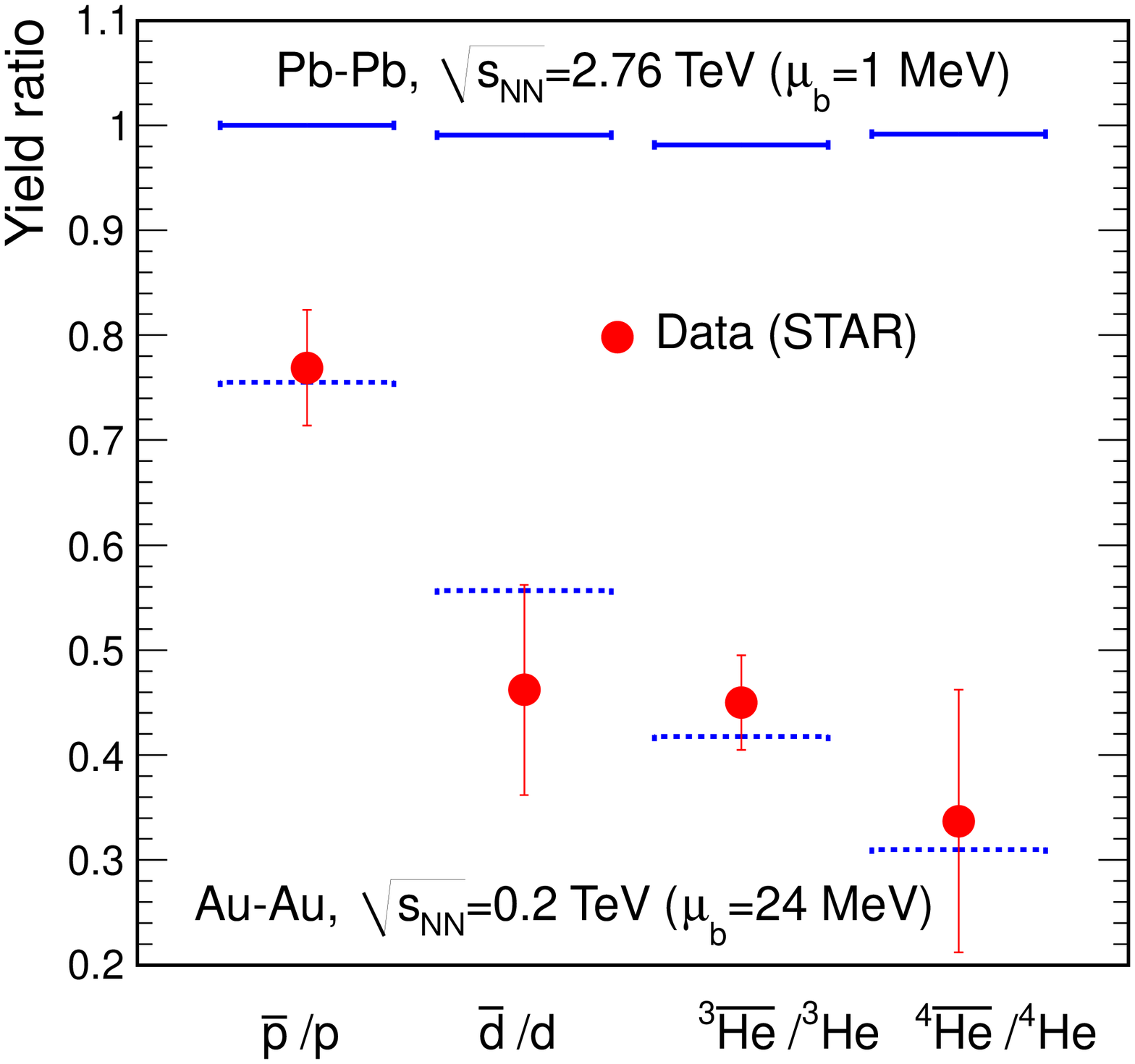}
\end{minipage}\end{tabular}
\caption{Model prediction for hadron yields relative to pions at LHC (left panel)
and of anti-matter to matter production at RHIC and LHC.}\label{fig_lhc}
\end{figure}

In Fig.~\ref{fig_lhc} we provide predictions for the  production of various
hadrons relative to pions, to be tested soon by experiment.
Preliminary ALICE data \cite{alice_1} indicate a lower $\bar{p}/\pi^-$ ratio. 
We show also how dramatic the balance between matter and anti-matter production 
is changing from RHIC to LHC energies, as illustrated by the ratios of 
anti-baryons to baryons. Prediction for (anti-)hyper-nuclei are also 
available \cite{aa10}.

We now turn to the heavy-quark sector and compare the model predictions to data on
charmonium prodution. 
Charmonium is considered, since the original proposal about its suppression in 
a Quark-Gluon Plasma (QGP) \cite{satz}, an important probe of the energy density
reached in the deconfined fireball produced in ultra-relativistic nucleus-nucleus
collisions (see \cite{satz_kluberg}).
Because the large mass of charm quarks, heavy flavor hadron production cannot be 
described in a purely thermal approach as that discussed above. 
It was realized in \cite{pbm1} that charmonium production can 
be well described in the statistical model by assuming that all charm quarks are 
produced in initial, hard collisions while charmed hadron and charmonium
production takes place exclusively at the phase boundary with statistical 
weights calculated in a thermal approach (for a recent review 
see \cite{pbm_js_lb}). 
An important element is thermal equilibration, at least near the critical 
temperature, $T_c$, which we believe can be achieved efficiently for charm 
only in the QGP. 
In recent publications \cite{aa2}  we have demonstrated that the data on 
J/$\psi$ and $\psi'$ production in nucleus-nucleus 
collisions at the SPS ($\sqrt{s_{NN}} \approx 17$ GeV) and RHIC 
($\sqrt{s_{NN}}$=200 GeV) energies can be well described within the statistical 
hadronization model.
We have recently shown \cite{aa09} that there are crucial differences
in elementary collisions compared to nucleus-nucleus for J/$\psi$ production.

Besides the thermal parameters discussed above, which we keep unchanged,
the model has as input parameter the charm production cross section in pp 
collisions,
used to calculate the number of directly produced $c\bar{c}$ pairs 
$N_{c\bar{c}}^{dir}$ which enter into the balance equation \cite{pbm1,pbm_js_lb}:
$N_{c\bar{c}}^{dir}=\frac{1}{2}g_c N_{oc}^{th}
\frac{I_1(g_cN_{oc}^{th})}{I_0(g_cN_{oc}^{th})} + g_c^2N_{c\bar c}^{th}$.

\begin{figure}[htb]
\begin{tabular}{cc}
\begin{minipage}{.49\textwidth}
\includegraphics[width=.95\textwidth]{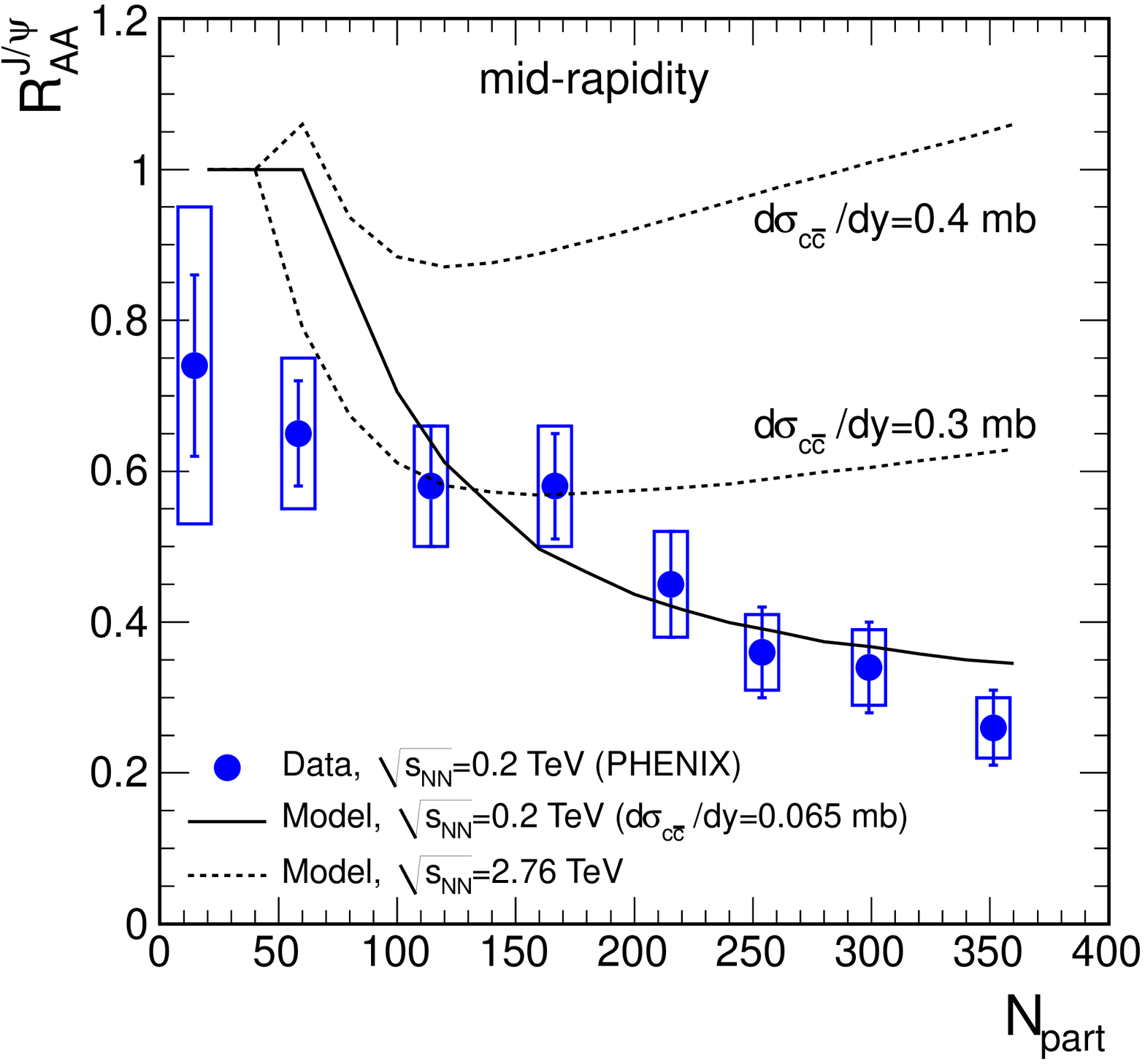}
\end{minipage}  & \begin{minipage}{.49\textwidth}
\includegraphics[width=.95\textwidth]{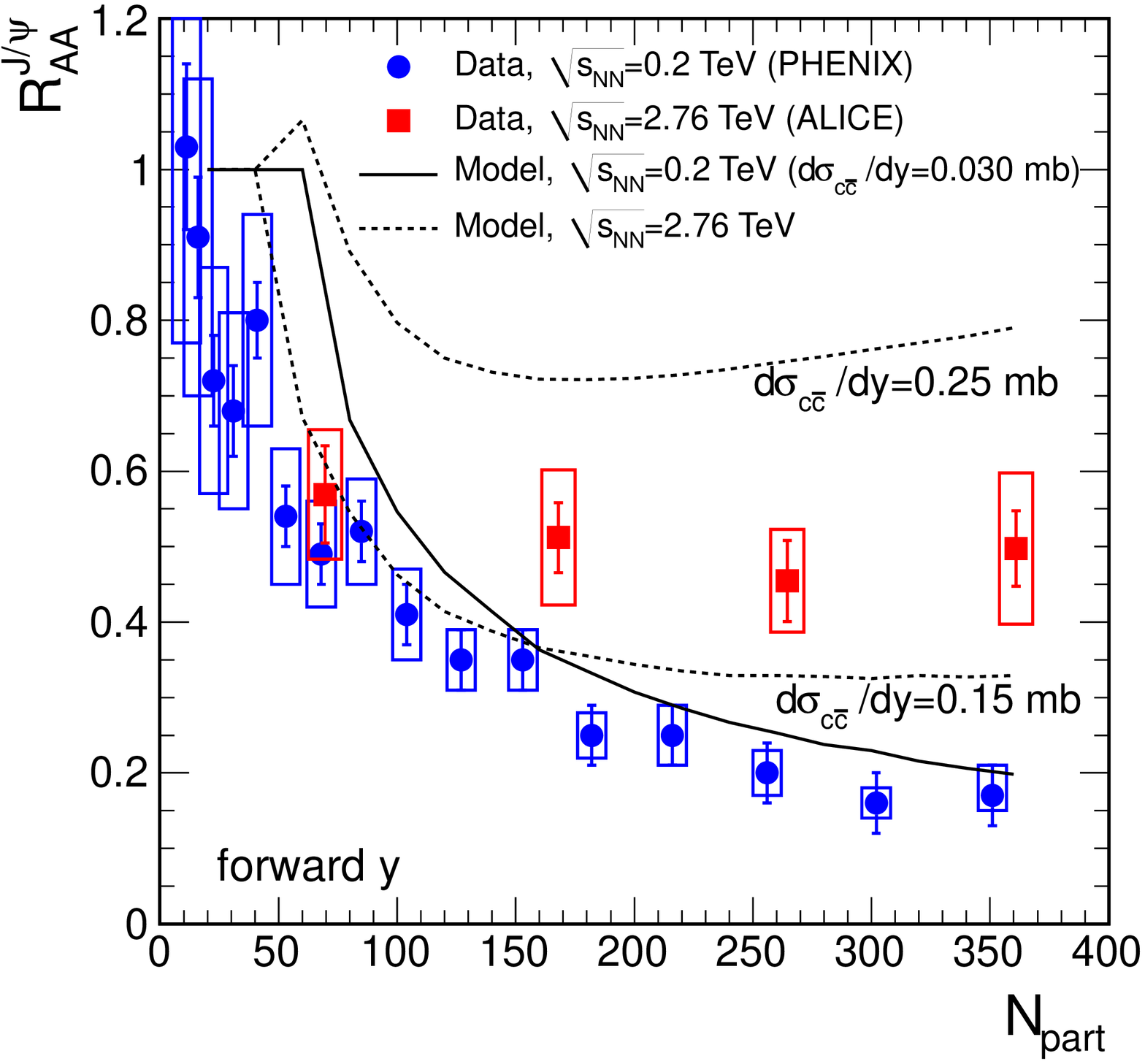}
\end{minipage}\end{tabular}
\caption{Centrality dependence of $R_{AA}^{J/\psi}$ for RHIC and LHC energies
at mid-rapidity (left panel) and forward rapidity (right panel). The two curves
shown for the LHC energy coorespond to a range of expected shadowing. 
The ALICE data shown in the right panel are preliminary results shown at this 
conference \cite{alice_2}.}
\label{fig_raa}
\end{figure}

The centrality dependence of the nuclear modification factor $R_{AA}^{J/\psi}$ 
is shown in Fig.~\ref{fig_raa}.   
The model reproduces very well the decreasing trend versus centrality seen 
in the RHIC data \cite{phe1}.
The larger $R_{AA}^{J/\psi}$ value at midrapidity is in our model due to the
enhanced generation of charmonium around mid-rapidity, determined by the
rapidity dependence of the charm production cross section.
At the much higher LHC energy the larger charm production cross section 
could lead to a different trend as a function of centrality, depending on the
magnitude of shadowing in Pb-Pb collisions. A generic prediction of the model is 
that the $R_{AA}^{J/\psi}$ value at LHC is larger than at RHIC and this is 
confirmed by the preliminary ALICE data \cite{alice_2} measured at forward 
rapidity, which demonstrate, in our view, that charmonium is produced at LHC 
at the phase boundary (chemical freeze-out).
If further confirmed  by data (importantly, also on $\psi'$ production), 
this picture will consolidate the role of charmonium 
as a special observable to probe deconfinement of heavy quarks and to 
delineate the phase boundary of QCD matter with hadrons carrying heavy quarks. 
At the LHC, first results of $\Upsilon$ production have appeared \cite{cms_ups}.
The model predicts for $\Upsilon$ a suppression-like pattern of a similar 
magnitude as that of J/$\psi$ at RHIC.
We predict \cite{aa09} for Pb-Pb much smaller ratios 
$\Upsilon(2S)/\Upsilon(1S)$=0.033 and $\Upsilon(2S)/\Upsilon(1S)$=0.005, 
compared to the values in $p\bar{p}$ collisions 
at Tevatron \cite{cdf_ups}, 0.32 and 0.15, respectively. 
This feature is indicated in the CMS data \cite{cms_ups}.

\end{document}